\documentclass[twocolumn,apl]{revtex4}
\usepackage{graphicx}
\usepackage{amsmath}
\usepackage{mathrsfs}
\usepackage{amsfonts}
\usepackage{amssymb}

\newcommand{\dpsi}{\mathrm{d}\psi}

\begin{document}

\title{Annular interdigital transducer focuses piezoelectric surface waves to a single point}

\author{Vincent Laude}
\affiliation{Institut FEMTO-ST, Universit\'e de Franche-Comt\'e, CNRS, F-25044 Besan\c{c}on, France}
\author{Davy G\'erard\footnote{D. G\'erard is now with the Institut Fresnel, Domaine Universitaire de Saint J\'{e}r\^{o}me, F-13397 Marseille cedex 20, France.}}
\affiliation{Institut FEMTO-ST, Universit\'e de Franche-Comt\'e, CNRS, F-25044 Besan\c{c}on, France}
\author{Naima Khelfaoui}
\affiliation{Institut FEMTO-ST, Universit\'e de Franche-Comt\'e, CNRS, F-25044 Besan\c{c}on, France}
\author{Carlos F. Jerez-Hanckes\footnote{C. F. Jerez-Hanckes is also with the Centre de Math\'ematiques Appliqu\'ees, CNRS, Ecole polytechnique, F-91128 Palaiseau, France, and EPCOS Sophia Design Center, 1300 Route des Cr\^etes, F-06560, Sophia-Antipolis, France.}}
\affiliation{Institut FEMTO-ST, Universit\'e de Franche-Comt\'e, CNRS, F-25044 Besan\c{c}on, France}
\author{Sarah Benchabane}
\affiliation{Institut FEMTO-ST, Universit\'e de Franche-Comt\'e, CNRS, F-25044 Besan\c{c}on, France}
\author{Abdelkrim Khelif}
\affiliation{Institut FEMTO-ST, Universit\'e de Franche-Comt\'e, CNRS, F-25044 Besan\c{c}on, France}


\begin{abstract}
%
We propose and demonstrate experimentally the concept of the annular interdigital transducer that focuses acoustic waves on the surface of a piezoelectric material to a single, diffraction-limited, spot.
The shape of the transducing fingers follows the wave surface.
Experiments conducted on lithium niobate substrates evidence that the generated surface waves converge to the center of the transducer, producing a spot that shows a large concentration of acoustic energy.
This concept is of practical significance to design new intense microacoustic sources, for instance for enhanced acouto-optical interactions.
\end{abstract}


\maketitle

A variety of anisotropic propagation phenomena are observed in solids, that are mainly determined by  the symmetries of the rank-four elastic constant tensor~\cite{roy1999}.
The piezoelectric effect offers a convenient means of generating elastic waves, by transforming electrical to mechanical energy.
The wide majority of elastic wave experiments in piezoelectric solids employ planar or interdigitated transducers (IDT) emitting plane waves with well-defined wave vectors~\cite{whi1965}.
In such a situation, a displacement field mimicking an ideal plane wave is often looked for and propagation is adequately described by phase quantities, such as the phase velocity.
In non piezoelectric materials, surface acoustic waves can be excited using a focused laser beam~\cite{von1983}.
In many instances, the surface can be considered to be excited at a single source point.
In this case, the surface supports outgoing waves that depart from the source and propagation in the far field depends on the angular dispersion of the spatial group velocity.
At a distance exceding a few wavelengths, the formed ripple pattern has the shape of the wave surface, obtained by plotting the group velocity as a function of the emission angle.
This effect was observed very neatly for surface waves in the experiments of Vines \textit{et al.}~\cite{vin1995} and Sugarawa \textit{et al.}~\cite{sug2002}.
Here, we investigate the converse problem: can we construct an extended source using an appropriate IDT that will focus elastic energy to a single point on the surface of a piezoelectric solid?

A circular IDT has already been proposed by Day and Koerber~\cite{day1972} for substrates with in-plane isotropy (orthotropy).
Since phase and group velocities are equal in this case, the wave surface reduces to a simple circle.
A more recent alternative, the so-called  focused interdigital transducer (FIDT), makes use of acoustic emission only within a circular arc~\cite{wu2005b,wan2006}.
In this case, the acoustic field is strongly dependent on the angular extent of the arc and on the focal length of the transducer.
Another limit comes from the anisotropy of the substrate which leads to aberrations at the focal point.
In the following, we consider using synchronous surface wave emission from all directions to obtain a diffraction limited focal point.


\begin{figure}[t]
 \centering
\includegraphics[width=80mm]{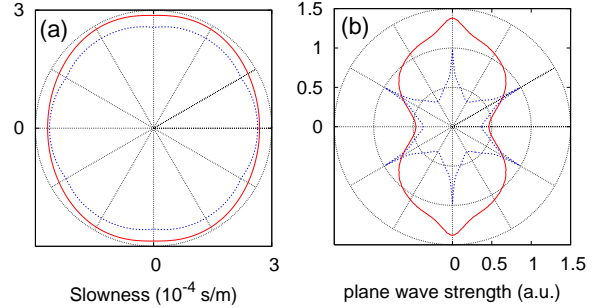}
\caption{[color online] Dependency of the anisotropic propagation of surface acoustic waves on the Y cut (solid red line) and the Z cut (dotted blue line) of lithium niobate (LiNbO$_3$): (a) slowness curve and (b) plane wave strength of the vertical surface displacement. In both plots, the horizontal axis is the crystallographic X axis.}
 \label{fig1}
\end{figure}

The anisotropic propagation of surface acoustic waves on a planar substrate is governed by two functions of the emission angle $\psi$.
The first one, $s(\psi)$, is the slowness (the inverse of the phase velocity).
The second function, $a(\psi)$, is a measure of the anisotropic plane wave strength; it is a number giving the rate of SAW emission in each direction~\cite{lau2006}.
Examples of slowness curves and  plane wave strengths are given for lithium niobate (LiNbO$_3$) in Figure~\ref{fig1}, for propagation on both the $(010)$ (Y~cut) and the $(001)$ (Z~cut) planes.
These functions depend acutely on the material cut considered and are clearly anisotropic.

\begin{figure}[t]
 \centering
\includegraphics[width=85mm]{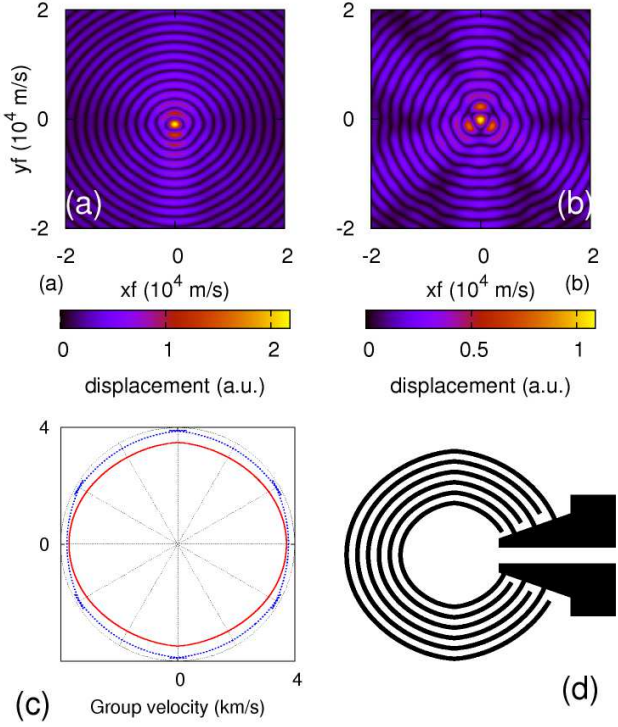}
\caption{[color online] Vertical displacement component of the surface Green's function for (a) Y cut and (b) Z cut lithium niobate. $x$ and $y$ are the coordinates of the observation point with respect to the source, and $f$ is the frequency. Wave surfaces for Y cut (solid red line) and Z cut (dotted blue line) lithium niobate are shown in (c). The principle of the annular interdigital transducer is depicted in (d).}
 \label{fig2}
\end{figure}

The displacement field of the outgoing wave originating from a point source is given by a Green's function with a free surface boundary condition.
The general form and a practical procedure to compute this Green's function were given in Ref.~\cite{lau2006}, for a time harmonic dependence of the form $\exp(\imath \omega t)$.
Of all possible acoustic contributions, only surface waves can efficiently transport energy in the far field.
Figures~\ref{fig2}a and~\ref{fig2}b display the modulus of the SAW contribution to the Green's function, with evanescent waves excluded, for the same orientations of lithium niobate as in Fig.~\ref{fig1}.
A system of ripples is seen to originate from the source center.
As the wave advances, the ripples become homothetic to the wave surface, as can be inferred from the following argument.
The SAW contribution to the far field of the Green's function can be estimated using the stationary phase principle~\cite{har2001}, with the result
\begin{equation}
\label{eq1}
G(R, \theta) \approx \frac{A a(\overline{\psi}) \exp{\left(-\imath \omega R h(\overline{\psi})\,-\,\imath\frac{\pi}{4}\mathrm{sgn}\left\{h''(\overline{\psi})\right\}\right)}}{\sqrt{\omega R\left|h''(\overline{\psi})\right|}}
\end{equation}
with $h(\psi) = s(\psi) \cos(\psi - \theta)$.
In this expression, the beam steering angle $\theta - \overline{\psi}$ is defined by the condition $\mathrm{d} h(\overline{\psi}) / \dpsi = h'(\overline{\psi}) = 0$; it is a function of the direction of observation $\theta$.
Note that the group velocity is $v_g(\theta) = h^{-1}(\overline{\psi})$.
$A$ is a constant independent of the position of observation.
Equation~(\ref{eq1}) indicates that in the far field, propagation occurs in the direction $\theta$ at the group velocity $v_g(\theta)$ and that the amplitude decreases with the square root of the distance.
The term $\sqrt{|h''(\overline{\psi})|}$ is proportional to the so-called phonon focusing factor, which gives cuspidal points and the caustics when it vanishes.
The wave surface, which is the locus of the group or energy velocity as a function of the emission direction~\cite{roy1999}, can be obtained directy from the slowness curve by numerical differentiation~\cite{lau2003a}.
Fig.~\ref{fig2}c displays the wave surface for both the Y and the Z cuts of lithium niobate.


\begin{figure}[t]
\centering
\includegraphics[width=75mm]{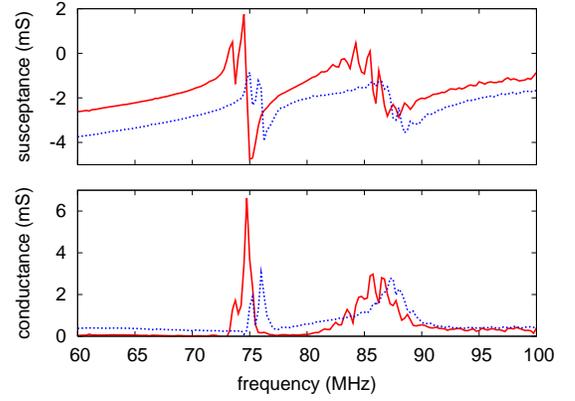}
\caption{[color online] Measured admittance of annular interdigital transducers on Y cut (solid red line) and Z cut (dotted blue line) lithium niobate.}
\label{fig3}
\end{figure}

We next define a pattern of electrodes with alternating electrical potentials such that the radiated SAW will be focused to a single point.
One possibility is to binarize the phase of the Green's function, thus producing a hologram of a source point.
Further considering that the source is in the far field of the focus point, we infer that the electrodes must be placed along a curve homothetic to the wave surface.
The design principle of such an annular interdigital transducer (AIDT) is depicted in Fig.~\ref{fig2}d, with the wave surface for Y cut lithium niobate used as an example.
Considering propagation along the direction given by angle $\theta$, it is well-known that the phase velocity vector is orthogonal to the wave surface~\cite{roy1999}.
By construction, the local pitch of the AIDT is then proportional to the phase velocity $s^{-1}(\overline{\psi}$).
Hence, the distance $p(\theta)$ between two electrodes can be chosen such that $f p(\theta) s(\overline{\psi}) = 1/2$.
This expression guarantees that the angular variation of $p(\theta)$ is such that the entire electrode array is at resonance at frequency $f$.
It also implies that the only wave that can be synchronous for all directions is the one selected when computing the wave surface.


Experiments have been conducted on Y and Z cut lithium niobate (LiNbO$_3$) substrates.
AIDT masks were designed such that the average acoustic wavelength $\lambda$ is 50~$\mu$m and the resonance frequency is 75~MHz.
The number of finger pairs is 200 and the metallization ratio is 0.5.
Figure~\ref{fig3} shows measurements of the admittance of AIDT devices for both cuts.
The response is roughly the superposition of a capacitance, the SAW contribution around 75~MHz, and a broad leaky-SAW contribution extending from 80 to 90~MHz approximately.
The SAW contribution has the form of two sharp peaks, one for the entrance and one for the exit of the stop band of the periodic electrode array.
In contrast, the leaky-SAW contribution is dispersed along the frequency axis since the emission of this type of wave is not synchronous.

\begin{figure}[t]
\centering
\includegraphics[width=85mm]{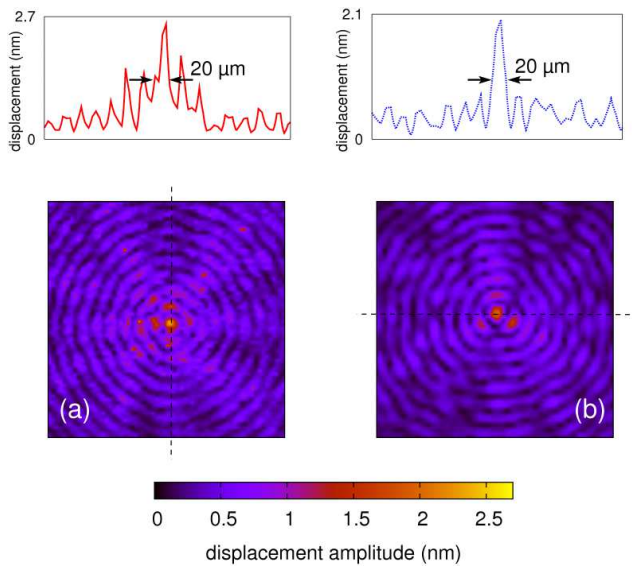}
\caption{[color online] Measured vertical displacements at the center of the annular interdigital transducers on (a) Y cut and (b) Z cut lithium niobate.
The scans are taken at the resonance frequency and cover an area of 400 by 400~$\mu$m$^2$. Cross sections through the focal point are shown above the scans.}
\label{fig4}
\end{figure}

To directly image and characterize the focal spot, the spatial distribution of the vertical displacement was measured thanks to an heterodyne optical probe~\cite{kok2007}.
The resulting maps are shown in Figure~\ref{fig4}.
Their spatial resolution, limited by the collection optics, is estimated to be 5~$\mu$m.
Waves emitted in all directions interfere in the central part of the device and form a concentric fringe pattern.
An important focusing in the center of the devices is clearly obtained at the main resonance frequency.
The maximum vertical displacement could be optimized by better electrical matching to the electrical source and further amplification.
We note, however, that local strain variations in the vicinity of the focus can be quite high.

Comparison between the experimental maps and the theoretical calculations of the acoustic field diffracted by a point source (Fig. \ref{fig2}) evidences that they bear a
striking resemblance.
This shows that the AIDT achieves the time reversal~\cite{ler2007} of the Dirac source considered for the Green's functions of Fig. \ref{fig2}.
Since the generated angular spectrum misses the evanescent waves required to reconstruct the near field, resolution is limited by diffraction.
For isotropic propagation on a surface, the amplitude at the focus would ideally be given by the Bessel function $J_0(kR)$, with $k=2\pi/\lambda$, resulting in a full width at half maximum (FWHM) of $0.48 \lambda$~\cite{hon2006}, or 24~$\mu$m for $\lambda = 50$~$\mu$m.
The experimental focus spot is slightly anisotropic, and its FWHM is approximately 20~$\mu$m for both the Y and the Z cut.
This value is slightly smaller than or comparable to the isotropic limit, though the difference is within the estimated spatial resolution of the scans.
This observation however calls for further studies of the diffraction limit for anisotropic propagation.


As a conclusion, we have proposed and demonstrated experimentally the concept of an annular interdigital transducer (AIDT) such that the fingers shape follows the wave surface.
The generated surface acoustic waves converge to the center of the transducer, producing an intense spot that is limited in resolution by diffraction only.
The AIDT can indeed be seen as a time-reversal device, reforming a diffraction-limited image of a point source that would be located at the resonator center.
This result is promising for several applications including intense microacoustic sources, e.g. for integrated acousto-optics, or to enhance non linear effects.

The authors wish to thank B. Wacogne and P. Vairac for their help with the AIDT masks and in setting up the heterodyne interferometer, respectively, and K. Kokkonen for insightful comments.


\end{document}